\documentclass[english,11pt,a4paper]{article}
\usepackage[T1]{fontenc}
\usepackage[latin9]{inputenc}
\usepackage{amsmath}
\usepackage{amsthm}
\usepackage{amssymb}
\usepackage{graphicx}
\usepackage{verbatim}
\makeatletter
\numberwithin{equation}{section}

\usepackage{jheppub}

\makeatother
\preprint{FERMILAB-PUB-23-0829-T}
\date{today}
\usepackage{babel}

\title{Imprints of light dark matter on the evolution of cosmic neutrinos}

\abstract{
Neutrinos are often considered as a portal to new physics beyond the Standard Model (SM) and might possess phenomenologically interesting interactions with dark matter (DM). 
This paper examines the cosmological imprints of DM that interacts with and is produced from SM neutrinos at temperatures below the MeV scale. 
We take a model-independent approach to compute the evolution of DM in this framework and present analytic results which agree well with numerical ones. Both freeze-in and freeze-out regimes are included in our analysis. 
Furthermore, we demonstrate that the thermal evolution of neutrinos might be substantially affected by their interaction with DM.
We highlight two distinctive imprints of such DM on neutrinos: (i) a large, negative contribution to $N_{\rm eff}$, which is close to the current experimental limits and will readily be probed by future experiments; (ii) spectral distortion of the cosmic neutrino background (C$\nu$B) due to DM annihilating into neutrinos, a potentially important effect for the ongoing experimental efforts to detect C$\nu$B.
}

\author[a,b]{Isaac R. Wang}
\author[c]{and Xun-Jie Xu}
\affiliation[a]{Theory Division, Fermi National Accelerator Laboratory, Batavia, IL 60510, USA}
\affiliation[b]{New High Energy Theory Center, Department of Physics and Astronomy, Rutgers University, Piscataway, NJ 08854, USA}
\affiliation[c]{Institute of High Energy Physics, Chinese Academy of Sciences, Beijing 100049, China}
\emailAdd{isaacw@fnal.gov}
\emailAdd{xuxj@ihep.ac.cn}

\begin{document}
\maketitle

\section{Introduction}
\label{sec:introduction}
Dark matter (DM) has been an enduring mystery for decades.
As extensive cosmological and astrophysical evidence supports its existence,
DM poses one of the major challenges to the success of the Standard Model (SM) of particle physics.
To date, DM searches in direct or indirect detection experiments, which rely on the couplings of DM with nucleons or electromagnetically-interacting particles, have not yet yielded any conclusive signals of DM.
Therefore, it is plausible to consider that DM may reside in a hidden dark sector that barely interacts with quarks or charged leptons.  For instance, the dark sector could be connected to the SM only via the neutrino portal~\cite{Falkowski:2009yz,Gonzalez-Macias:2016vxy,Escudero:2016tzx,Escudero:2016ksa,Batell:2017rol,Becker:2018rve,Blennow:2019fhy,Hufnagel:2021pso,Barman:2022scg,Li:2022bpp,Xu:2023xva}, suggesting a scenario where DM predominantly interacts with neutrinos rather than other SM particles.
This scenario, despite being generically difficult to probe in direct or indirect detection experiments, may still leave discernible cosmological imprints.

Indeed, as already investigated in numerous studies, the presence of DM-neutrino interactions may cause cosmologically observable effects.
These include modifying the cosmic microwave background (CMB) anisotropies~\cite{Serra:2009uu,Wilkinson:2014ksa,Becker:2020hzj,Mosbech:2020ahp},  altering the big bang nucleosynthesis (BBN) predictions~\cite{Depta:2019lbe,Sabti:2019mhn}, reducing small-scale structures~\cite{Mangano:2006mp,Bertoni:2014mva}, etc.

In this work, we focus on the epoch when the SM neutrinos ($\nu$) have decoupled from the thermal bath.
We examine a generic framework where a light DM candidate, denoted by $\chi$, interacts predominantly with $\nu$. The interaction leads to the conversion $\nu\overline{\nu}\leftrightarrow\chi\overline{\chi}$ and is responsible for the DM abundance, via either the freeze-in~\cite{Hall:2009bx} or the freeze-out~\cite{Kolb:1990vq} mechanism.

Due to the conversion between $\chi$ and $\nu$, the effective number of neutrino species, $N_{\rm eff}$, could receive a potentially large correction, which can be  positive~\cite{Berlin:2017ftj,Berlin:2018ztp} or negative~\cite{Hufnagel:2021pso}, depending on the strength and energy scale of the interaction. While a positive correction to $N_{\rm eff}$ is rather common in various new physics scenarios (see e.g.~Ref.~\cite{Boehm:2012gr, Huang:2017egl,Abazajian:2019oqj,Luo:2020fdt,Luo:2020sho, Yeh:2022heq,Li:2022dkc,Li:2023puz,Ghosh:2023ocl}), a negative one is relatively rare. In Ref.~\cite{Hufnagel:2021pso}, the negative correction is obtained only in the freeze-in regime but in fact this can also occur in the freeze-out regime, as we will show in this work.

The impact of such DM is not limited to $N_{\rm eff}$; it also extends to the energy distribution of cosmic neutrinos.
The process $\chi\overline{\chi}\to\nu\overline{\nu}$ continues to generate neutrinos until today at a low, yet non-vanishing rate after its decoupling.
This would distort the high-energy tail of the cosmic neutrino background (C$\nu$B) from the exponentially suppressed form to a power-law form, and hence substantially enhance the neutrino flux of C$\nu$B at high energies. In particular, we will show that the enhanced C$\nu$B tail yields a neutrino flux much higher than that from BBN isotope decays and solar thermal production in the sub-eV to keV range.

This paper is organized as follows. In Sec.~\ref{sec:evolution}, we calculate the thermal evolution of DM and neutrinos in the presence of DM-neutrino interactions.
In Sec.~\ref{sec:consequence}, we evaluate the impact of DM on $N_{\rm eff}$ and C$\nu$B distortion, as well as the Lyman-$\alpha$ constraints on light DM.
Finally we conclude in Sec.~\ref{sec:summary} and relegate some details to appendices.

\section{Evolution of the dark sector}
\label{sec:evolution}

\begin{figure}
	\centering

	\includegraphics[width=0.8\textwidth]{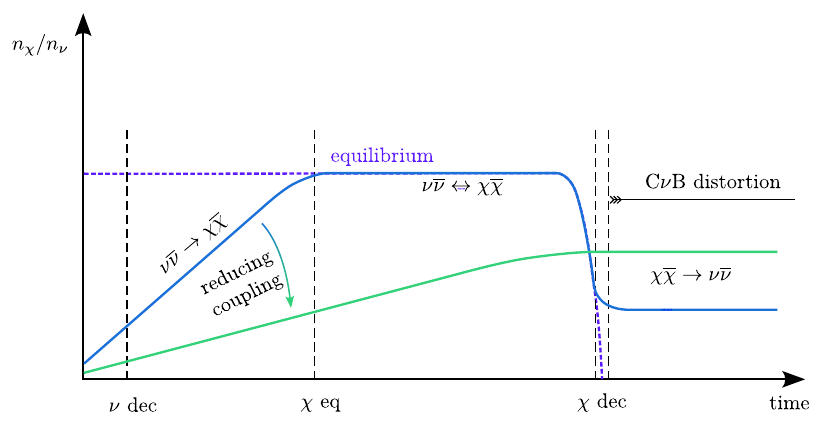}\caption{A schematic showing the evolution of the DM abundance in our framework.
		Here ``$\nu$ dec'', ``$\chi$ eq'', and ``$\chi$ dec'' indicate
		the moments of neutrino decoupling, $\chi$ reaching equilibrium,
		and $\chi$ decoupling. The period between ``$\chi$ eq'', and ``$\chi$
		dec'' may be absent for scenarios with sufficiently weak interactions,
		such as the green curve. For ``$\chi$ dec'', we plot two vertical
		dashed lines to represent the difference between chemical and kinetic
		decoupling. C$\nu$B distortion due to $\chi\overline{\chi}\to\nu\overline{\nu}$
		only starts from the kinetic decoupling.  \label{fig:schematic}
	}
\end{figure}

\subsection{A schematic of the framework}

Figure~\ref{fig:schematic} illustrates schematically the evolution
of DM abundance in our framework. In the early universe when
the temperature $T$ is well above all mass scales involved in $\nu\overline{\nu}\leftrightarrow\chi\overline{\chi}$,
the reaction rate $\Gamma\propto T$ can be inefficient in comparison
to the Hubble expansion rate $H\propto T^{2}$. Within this epoch,
$\chi$ is not in equilibrium with $\nu$, and the number of $\chi$
particles in a comoving volume keeps increasing due to $\nu\overline{\nu}\to\chi\overline{\chi}$ towards its equilibrium value.
We refer to this epoch as the production phase of DM.

Depending on the strength of the interaction, the subsequent evolution may or may not
be able to drive $\chi$ into equilibrium, as demonstrated by the blue and green
curves in Fig.~\ref{fig:schematic}, respectively.
In the former case,
the DM number density, $n_\chi$, is kept at its equilibrium value until $n_\chi$ is too low to maintain thermal equilibrium, leading to the well-known freeze-out mechanism.
While for the latter case, $\chi$ is produced at a low reaction rate which is always below the Hubble expansion rate, corresponding to the freeze-in mechanism.

After $\chi$ has decoupled with $\nu$, the process  $\chi\overline{\chi}\to\nu\overline{\nu}$ may lead to C$\nu$B distortion.
Note that there is a difference between chemical decoupling
and kinetic decoupling. The chemical equilibrium is maintained by $\nu\overline{\nu}\leftrightarrow\chi\overline{\chi}$
which can alter particle numbers of $\nu$ or $\chi$, while the
kinetic equilibrium is maintained by processes that can transfer the kinetic energy
of $\nu$ and $\chi$ from one to the other. Hence the kinetic equilibrium
can be maintained by not only $\nu\overline{\nu}\leftrightarrow\chi\overline{\chi}$
but also the particle-number-conserving process $\nu\chi\leftrightarrow\nu\chi$.
Consequently, kinetic equilibrium typically lasts longer than chemical
equilibrium, as we will see in a specific example in Sec.~\ref{sec:CnuB}.
As long as the kinetic equilibrium is maintained, it is justified
to assume
\begin{equation}
	f_{\nu}(p)\propto f_{\nu}^{{\rm eq}}(p)\thinspace,\ \ f_{\chi}(p)\propto f_{\chi}^{{\rm eq}}(p)\thinspace,\label{eq:-20}
\end{equation}
where $f_{\nu/\chi}$ denotes the phase space distribution function
of $\nu/\chi$, and $f_{\nu,\chi}^{{\rm eq}}$ denotes the equilibrium
value of $f_{\nu/\chi}$ when both kinetic equilibrium and chemical
equilibrium are reached. After kinetic decoupling, however, $f_{\nu}$
and $f_{\chi}$ are unable to maintain the shape in Eq.~\eqref{eq:-20}.
In particular, due to DM annihilation, $\chi\overline{\chi}\to\nu\overline{\nu}$,
which keeps generating neutrinos until today at a very low yet non-vanishing
rate, we expect that today's $f_{\nu}$ should be distorted. This
is indicated in Fig.~\ref{fig:schematic} as C$\nu$B distortion.

In what follows, we will study quantitatively the evolution depicted in Fig.~\ref{fig:schematic}, and scrutinize these effects on cosmic neutrinos.

Before we start, we shall clarify our notation of temperature in this work.
Since neutrinos and photons have different temperatures after $e^{+}e^{-}$
annihilation, we denote them by $T_{\nu}$ and $T_{\gamma}$ when
they need to be distinguished. Otherwise, we use $T$ to denote a
generic temperature. More strictly, we define $T$ as
\begin{equation}
	T\equiv T_{{\rm \star}}\frac{a_{\star}}{a}\,,
\end{equation}
where $T_{\star}$ denotes a pivot temperature that can be set at
any point before neutrino decoupling (e.g.~$T_{\star}=10$ MeV), and $a$ is the scale factor
with ``$\star$'' indicating the pivot value.

\subsection{The Boltzmann equation}

\begin{figure}
	\centering

	\includegraphics[width=0.7\textwidth]{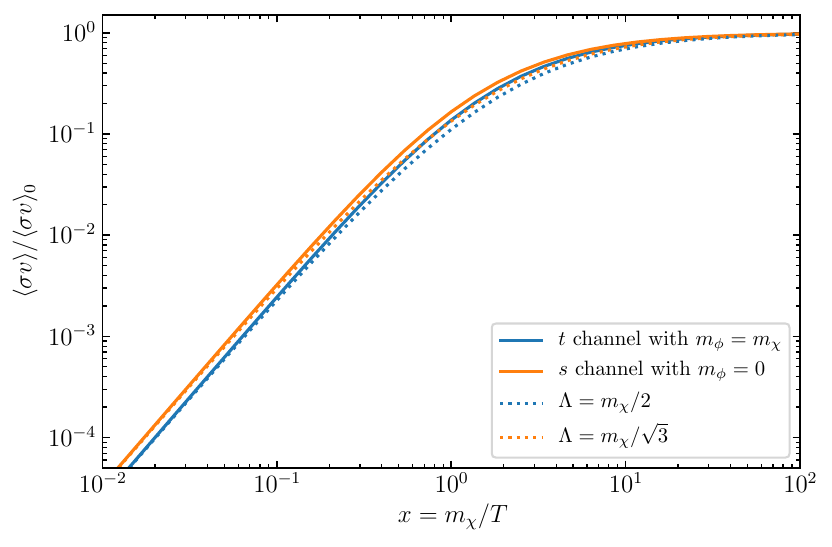}\caption{The temperature dependence of $\langle\sigma v\rangle$. The two solid
		curves represent two possible UV completions of the process $2\nu\leftrightarrow2\chi$
		through an $s$- or $t$-channel mediator $\phi$---see Eqs.~\eqref{eq:s}
		and \eqref{eq:t}. The dotted curves represent our phenomenological
		modeling of $\langle\sigma v\rangle$ given in Eq.~\eqref{eq:sigmav-formal},
		with the free parameter $\Lambda$ determined by the high-$T$ limits
		of the solid curves. \label{fig:sigma-v}}
\end{figure}

We start with a  model-independent analysis by considering  the thermally averaged
cross section, $\langle\sigma v\rangle$, which has been extensively used
in DM studies, in particular, in non-relativistic freeze-out scenarios. We should mention here that $\langle\sigma v\rangle$ can be rigorously
defined also in the relativistic regime~\cite{Gondolo:1990dk}, making it applicable to both
freeze-in and freeze-out scenarios.
Moreover, we will show that
it can be parametrized by two simple parameters which suffice for
an accurate description of the collision terms in specific models
(see Fig.~\ref{fig:sigma-v}). Complete analyses for specific models
are also conducted in this work, arriving at
similar results (see Sec.~\ref{subsec:Model-specific-analyses} and Fig.~\ref{fig:DM-abundance}).

Given the thermally averaged cross section for  $\nu\overline{\nu}\leftrightarrow\chi\overline{\chi}$,
the number density of DM is governed by the following Boltzmann equation:
\begin{equation}
	\frac{dn_{\chi}}{dt}+3Hn_{\chi}=-N_{\nu}\langle\sigma v\rangle\left(n_{\chi}^{2}-n_{\nu}^{2}B^{2}\right),\label{eq:-8}
\end{equation}
where $n_{\chi}$ and $n_{\nu}$ denote the number densities of $\chi$
and $\nu$, $H$ is the Hubble parameter, $N_{\nu}=3$ accounts for
the three neutrino flavors, and the Boltzmann suppression factor $B$
is defined as
\begin{equation}
	B\equiv\frac{x^{2}}{2}K_{2}\left(x\right),~~x\equiv \frac{m_{\chi}}{T}\,,
	\label{eq:-9}
\end{equation}
where $m_{\chi}$ is the mass of $\chi$.
Eq.~\eqref{eq:-8} together with the $B$ factor in
Eq.~\eqref{eq:-9} can be obtained by reformulating the conventional
collision terms into the $\langle\sigma v\rangle$ form, assuming
the validity of Eq.~\eqref{eq:-20} and Boltzmann statistics. In Appendix~\ref{sec:Collision-terms},
we briefly review the derivation of Eq.~\eqref{eq:-8}.

Note that in our convention, $n_{\nu}$ is defined as the neutrino
number density of a single flavor, without including antineutrinos.
The same convention also applies to $n_{\chi}$. In this work, we
assume that DM-neutrino interactions are flavor-universal and flavor-diagonal.

When $\nu$ is converted to $\chi$ via $\nu\overline{\nu}\leftrightarrow\chi\overline{\chi}$,
the total number of $\nu$ and $\chi$ in a comoving volume is conserved
if they do not interact with other particles. Hence, we define
\begin{equation}
	n_{\chi\nu}\equiv n_{\chi}+N_{\nu}n_{\nu}\thinspace,\ \overline{n}_{\chi\nu}\equiv\overline{n}_{\chi}+N_{\nu}\overline{n}_{\nu}\thinspace,\label{eq:-21}
\end{equation}
where $\overline{n}_{\chi}\equiv n_{\chi}a^{3}$ and $\overline{n}_{\nu}\equiv n_{\nu}a^{3}$,
with $a$ the scale factor. After neutrino decoupling, $\overline{n}_{\chi\nu}$
should be a constant, $d\overline{n}_{\chi\nu}/dt=0$.

The thermally averaged cross section $\langle\sigma v\rangle$ is
defined as
\begin{equation}
	\langle\sigma v\rangle\equiv n_{\chi}^{-2}\int f_{\chi}(p_{1})f_{\chi}(p_{2})|{\cal M}|^{2}(2\pi\delta)^{4}d\Pi_{1}d\Pi_{2}d\Pi_{3}d\Pi_{4}\thinspace,\label{eq:-10}
\end{equation}
where $d\Pi_{i}\equiv\frac{d^{3}p_{i}}{(2\pi)^{3}2E_{i}}$ with $i$
indicating the $i$-th particle in $\chi\overline{\chi}\to\nu\overline{\nu}$,
$|{\cal M}|^{2}$ denotes the squared amplitude of the reaction, and
$(2\pi\delta)^{4}\equiv(2\pi)^{4}\delta^{4}(p_{1}+p_{2}-p_{3}-p_{4})$.
Despite its common use in non-relativistic freeze-out, $\langle\sigma v\rangle$
is well defined in the relativistic regime as well. So Eq.~\eqref{eq:-10}
can be readily applied to both non-relativistic and relativistic annihilation.

Let us investigate the behavior of $\langle\sigma v\rangle$ at different
temperatures. In the low-$T$ limit when $\chi$ becomes non-relativistic,
$\langle\sigma v\rangle$ can be regarded as a constant, $\langle\sigma v\rangle=\langle\sigma v\rangle_{0}$,
provided that $\chi\overline{\chi}\to\nu\overline{\nu}$ is $s$-wave
annihilation. In the high-$T$ limit when all masses are well below
$T$ and hence negligible, we expect from a simple dimensional analysis
that
\begin{equation}
	\langle\sigma v\rangle\propto T^{-2}\thinspace.\label{eq:-11}
\end{equation}

Therefore, for practical use,
we parametrize
$\langle\sigma v\rangle$ as
\begin{equation}
	\langle\sigma v\rangle\approx\frac{\langle\sigma v\rangle_{0}}{\left(1+T/\Lambda\right)^{2}}\thinspace,\label{eq:sigmav-formal}
\end{equation}
where $\Lambda$ represents the energy scale (typically the highest
mass of those involved  in the annihilation processes) when $\langle\sigma v\rangle$
starts to transition from its constant value $\langle\sigma v\rangle_{0}$
to the $T^{-2}$ form.

Eq.~\eqref{eq:sigmav-formal} provides a simple parametrization of $\langle\sigma v\rangle$
with only two free parameters,  $\langle\sigma v\rangle_{0}$ and
$\Lambda$. Despite its simplicity, Eq.~\eqref{eq:sigmav-formal}
can accurately account for the temperature dependence of $\langle\sigma v\rangle$ in a
specific model.

For illustration, let us consider two models where $\nu\overline{\nu}\leftrightarrow\chi\overline{\chi}$
is mediated by either an $s$-channel vector mediator $\phi^{\mu}$
or a $t$-channel scalar mediator $\phi$:
\begin{align}
	s\text{-channel model}:\ \ {\cal L} & \supset g_{\nu}\nu^{\dagger}\overline{\sigma}^{\mu}\phi{}_{\mu}\nu+g_{\chi}\chi^{\dagger}\overline{\sigma}^{\mu}\phi{}_{\mu}\chi\thinspace,\label{eq:s} \\
	t\text{-channel model}:\ \ {\cal L} & \supset y_{\chi}\nu\chi\phi\thinspace,\label{eq:t}
\end{align}
where $g_{\nu/\chi}$ and $y_{\chi}$ are dimensionless couplings.
We denote the mediator mass by $m_{\phi}$. It should be above
$m_{\chi}$ in the $t$-channel model as required by the stability
of $\chi$, while in the $s$-channel model, it can be arbitrarily light.

For the above specific models, the  thermally-averaged cross sections are computed in Appendix~\ref{sec:Collision-terms}.
The numerical results of $\langle\sigma v\rangle$ for the two models are presented in Fig.~\ref{fig:sigma-v} by the solid curves.
In certain
limits,
they have analytic forms:
\begin{align}
	\langle\sigma v\rangle^{(t)}\approx\frac{y_{\chi}^{4}}{32\pi m_{\chi}^{2}}\times\begin{cases}
		                                                                                1                & \text{for }x\gg1 \\
		                                                                                \frac{1}{4}x^{2} & \text{for }x\ll1
	                                                                                \end{cases}\thinspace, & \ \ \langle\sigma v\rangle^{(s)}\approx\frac{g_{\nu}^{2}g_{\chi}^{2}}{32\pi m_{\chi}^{2}}\times\begin{cases}
		                                                                                                                                                                                                        1                & \text{for }x\gg1 \\
		                                                                                                                                                                                                        \frac{1}{3}x^{2} & \text{for }x\ll1
	                                                                                                                                                                                                        \end{cases}\thinspace,\label{eq:-7}
\end{align}
where the superscripts $(t)$ and $(s)$ indicate the type of the
mediator. By fitting Eq.~\eqref{eq:sigmav-formal} to  Eq.~\eqref{eq:-7},
we obtain $\Lambda=m_{\chi}/2$ and $\Lambda=m_{\chi}/\sqrt{3}$ for
$\langle\sigma v\rangle^{(t)}$ and $\langle\sigma v\rangle^{(s)}$,
respectively. As is shown in Fig.~\ref{fig:sigma-v}, the dotted curves produced from Eq.~\eqref{eq:sigmav-formal} fit the solid curves rather accurately.

\subsection{Analytic solution in the high-$T$ regime}

In the high-$T$ regime, Eq.~\eqref{eq:sigmav-formal} gives $\langle\sigma v\rangle\approx\langle\sigma v\rangle_{0}\Lambda^{2}/T^{2}$.
In addition, the Boltzmann suppression factor $B$ in Eq.~\eqref{eq:-8}
can be neglected, $B\approx1$. In this regime, the Boltzmann equation
can be solved analytically, as can be seen by rewriting Eq.~\eqref{eq:-8}
as follows:
\begin{equation}
	\frac{d\overline{n}_{\chi}}{da}=N_{\nu}\frac{\langle\sigma v\rangle_{0}}{H_{\Lambda}a_{\Lambda}^{4}}\left(\overline{n}_{\nu}^{2}-\overline{n}_{\chi}^{2}\right),\label{eq:-12}
\end{equation}
where $a_{\Lambda}$ and $H_{\Lambda}$ denote the values of $a$
and $H$ at $T=\Lambda$.
Using the conservation of $\overline{n}_{\chi\nu}$
introduced in Eq.~\eqref{eq:-21} and substituting $\overline{n}_{\nu}=(\overline{n}_{\chi\nu}-\overline{n}_{\chi})/N_{\nu}$
into Eq.~\eqref{eq:-12}, we obtain the following analytic solution:
\begin{equation}
	\overline{n}_{\chi}=\overline{n}_{\chi\nu}\frac{{\cal E}-1}{4{\cal E}+2}\thinspace,\label{eq:-13}
\end{equation}
with
\begin{equation}
	{\cal E}=\exp\left(2R_{\Lambda}\frac{a}{a_{\Lambda}}\right),\ \ R_{\Lambda}\equiv\frac{\langle\sigma v\rangle_{0}\overline{n}_{\chi\nu}}{H_{\Lambda}a_{\Lambda}^{3}}=\left.\frac{\langle\sigma v\rangle_{0}\text{\ensuremath{n_{\chi\nu}}}}{H}\right|_{T\to\Lambda}.\label{eq:-14}
\end{equation}
Here we have already taken $N_{\nu}=3$. For a more general value
of $N_{\nu}$, $4{\cal E}+2$ in Eq.~\eqref{eq:-13} should be replaced
by $(N_{\nu}+1){\cal E}+N_{\nu}-1$.

The dimensionless ratio $R_{\Lambda}$ in Eq.~\eqref{eq:-14} quantifies
the rapidity of $\overline{n}_{\chi}$ approaching its equilibrium
value. For $R_{\Lambda}\gg1$, we expect that ${\cal E}\gg1$ at $a\gtrsim a_{\Lambda}/R_{\Lambda}$.
Consequently, Eq.~\eqref{eq:-13} gives $\overline{n}_{\chi}=\overline{n}_{\chi\nu}/4$,
implying that $\chi$ reaches the same number density as $\nu$ (per
flavor). For $R_{\Lambda}\ll1$, $\chi$ cannot reach equilibrium
before $a=a_{\Lambda}$ while the subsequent evolution would further
reduce $\langle\sigma v\rangle\text{\ensuremath{n_{\chi\nu}}}/H$
because at $T\ll\Lambda$ we have $\langle\sigma v\rangle n_{\chi\nu}\propto a^{-3}$
and $H\propto a^{-2}$. Therefore, $R_{\Lambda}\gtrsim1$ can be used
as the criteria to infer whether $\chi$ is able to reach equilibrium.

In the left panel of Fig.~\ref{fig:sol}, we plot Eq.~\eqref{eq:-13}
by the red dashed curve for $m_{\chi}=10$ keV, $\Lambda=m_{\chi}/\sqrt{3}$,
and $\langle\sigma v\rangle_{0}=1.6\times10^{-21}m_{\chi}^{2}$, corresponding
to $R_{\Lambda}=56.8$. The curve agrees well with the numerical solution
in the high-$T$ regime.

\subsection{Analytic solution in the low-$T$ regime}

In the low-$T$ regime when $T\ll\Lambda$, Eq.~\eqref{eq:sigmav-formal}
gives $\langle\sigma v\rangle\approx\langle\sigma v\rangle_{0}$.
By defining the following variables
\begin{equation}
	y=\frac{N_{\nu}\langle\sigma v\rangle_{0}}{H_{m}a_{m}^{3}}\overline{n}_{\chi}\thinspace,\ \ y_{0}=\frac{N_{\nu}\langle\sigma v\rangle_{0}}{H_{m}a_{m}^{3}}\overline{n}_{\nu}B\thinspace,\ x\equiv\frac{m_{\chi}}{T}\thinspace,\label{eq:-15}
\end{equation}
where $a_{m}$ and $H_{m}$ denote the values of $a$ and $H$ at
$T=m_{\chi}$, we can rewrite Eq.~\eqref{eq:-8} into the  form of
a Riccati equation:
\begin{equation}
	y'(x)=-\frac{y(x)^{2}-y_{0}(x)^{2}}{x^{2}}\thinspace.\label{eq:-16}
\end{equation}
In general, Eq.~\eqref{eq:-16} cannot be solved analytically.
But when $y_{0}$ is negligibly small (it is Boltzmann suppressed at  $x\gg1$),
Eq.~\eqref{eq:-16} can be solved analytically in the limit of vanishing $y_0$.
Following the computation in Appendix~\ref{sec:f-o-approx}, we find that the following piecewise function can approximate the exact solution quite accurately:
\begin{equation}
	y=\begin{cases}
		\frac{x}{x/x_{{\rm f.o}}-1}                                     & x>2x_{{\rm f.o}}                     \\
		y_{0}(x)                                                        & x<x_{{\rm f.o}}                      \\
		\frac{x\thinspace y_{{\rm f.o}}}{x(\xi-1)+x_{{\rm f.o}}(2-\xi)} & x\in[x_{{\rm f.o}},\ 2x_{{\rm f.o}}]
	\end{cases}\thinspace,\label{eq:-4-1}
\end{equation}
where $y_{{\rm f.o}}\equiv y_{0}(x_{{\rm f.o}})$ and $\xi\equiv y_{{\rm f.o}}/x_{{\rm f.o}}$
with $x_{{\rm f.o}}$ the freeze-out value of $x$. Note that Eq.~\eqref{eq:-4-1} is constructed by continuously connecting the known solutions for $x\gg 2x_{\rm f.o}$ and  $x\ll x_{\rm f.o}$ but such a construction cannot guarantee the continuity of the first-order derivative. Consequently, Eq.~\eqref{eq:-4-1} may exhibit slight non-smoothness in the transition range between the two regimes.
In Appendix~\ref{sec:f-o-approx}, we find that $x_{\rm f.o}$ can be approximately evaluated by
\begin{align}
	\label{eq:xfo}
	x_{\rm f.o} \simeq 0.048 + 1.73 \log_{10}(y_{0i}) + 0.051 (\log_{10}(y_{0i}))^2,
\end{align}
where $y_{0i}$ is the initial value of $y_0$, under the assumption that the equilibrium of $\nu \bar{\nu} \leftrightarrow \chi \bar{\chi}$ is maintained. More specifically,
\begin{align}
	\label{eq:-17}
	y_{0i} = \lim_{x \to 0} y_0(x) = \frac{y_0(x)}{B(x)}\,.
\end{align}

\begin{figure}
	\centering

	\includegraphics[width=0.49\textwidth]{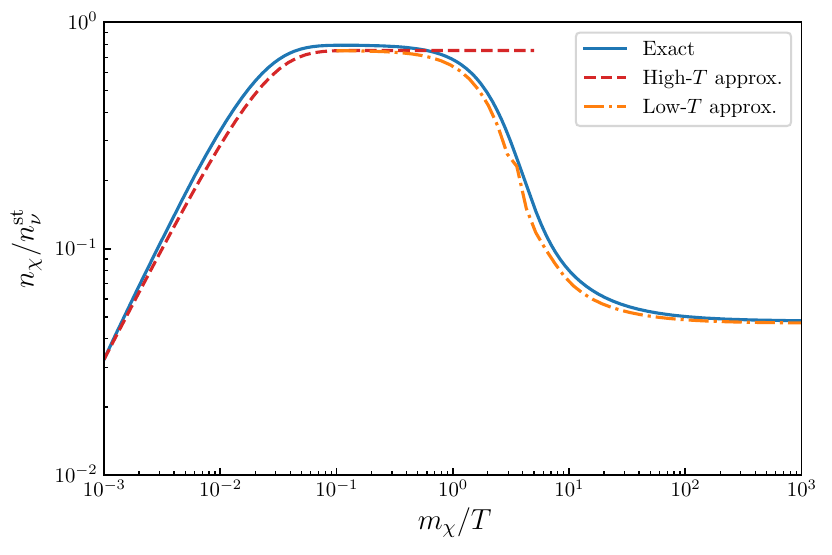}\includegraphics[width=0.495\textwidth]{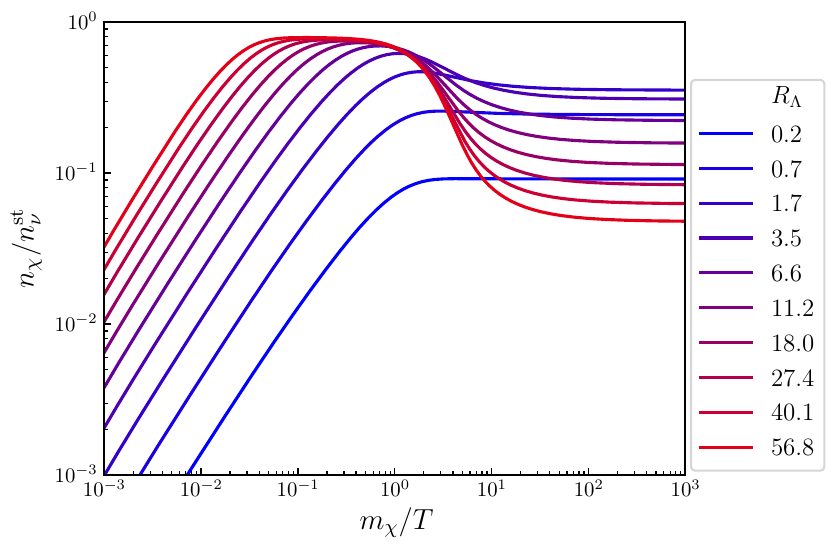}\caption{Left panel: analytic solutions obtained in the high-$T$ (dashed curve)
	and low-$T$ (dash-dotted) approximations compared with the numerical
	one (blue). Right panel: numerical solutions for $R_{\Lambda}$ varying
	from the weakly-coupled regime ($R_{\Lambda}\ll1$) to the strongly-coupled
	regime ($R_{\Lambda}\gg1$). The $y$-axes have been normalized by
	$n_{\nu}^{{\rm st}}$ which is the value of $n_{\nu}$ in the standard
	cosmological model. \label{fig:sol}}
\end{figure}

Combining Eq.~\eqref{eq:-17} with Eq.~\eqref{eq:-15}, we obtain
$y/y_{0i}=\overline{n}_{\chi}/\overline{n}_{\nu}$.
Here $\overline{n}_{\nu}$
could be treated as a constant but the approximation can be further
improved by taking into account the increase of $\overline{n}_{\nu}$
due to DM annihilation. More specifically, when $\chi$ is in equilibrium
with $\nu$, we have $\overline{n}_{\chi}=\overline{n}_{\nu}B$. Combining
with the conservation of particle number, $N_{\nu}\overline{n}_{\nu}+\overline{n}_{\chi}=N_{\nu}\overline{n}_{\nu}^{{\rm st}}$
where $\overline{n}_{\nu}^{{\rm st}}$ stands for the value of $\overline{n}_{\nu}$
in the standard cosmological model, we obtain $\overline{n}_{\nu}=\frac{N_{\nu}}{N_{\nu}+B}\overline{n}_{\nu}^{{\rm st}}$.
As $B$ decreases from $1$ to a vanishing value, $\overline{n}_{\nu}$
increases from $\overline{n}_{\nu}\approx\frac{3}{4}\overline{n}_{\nu}^{{\rm st}}$
to $\overline{n}_{\nu}\approx\overline{n}_{\nu}^{{\rm st}}$. In
this way, we obtain the following result:
\begin{equation}
	\overline{n}_{\chi}=\frac{y(x)}{y_{0i}}\frac{N_{\nu}}{N_{\nu}+B(x)}\overline{n}_{\nu}^{{\rm st}}\thinspace,\label{eq:-18}
\end{equation}
where $y(x)$ takes the piecewise function in Eq.~\eqref{eq:-4-1}.

In the left panel of Fig.~\ref{fig:sol}, we plot Eq.~\eqref{eq:-18}
by the orange dash-dotted curve for $m_{\chi}=10$ keV, $\Lambda=m_{\chi}/\sqrt{3}$,
and $\langle\sigma v\rangle_{0}=1.6\times10^{-21}m_{\chi}^{2}$, corresponding
to $R_{\Lambda}=56.8$. The curve agrees well with the numerical solution
in the low-$T$ regime.

In the right panel of Fig.~\ref{fig:sol}, we vary $R_{\Lambda}\propto\langle\sigma v\rangle_{0}$
from the strongly-coupled regime ($R_{\Lambda}\gg1$) to the weakly-coupled
regime ($R_{\Lambda}\ll1$) and plot the numerical solutions. The
plot shows that, as $R_{\Lambda}$ increases, the resulting DM abundance
first increases until $R_{\Lambda} \simeq 1$ and then decreases.

\subsection{DM relic abundance}

\begin{figure}
	\centering

	\includegraphics[width=0.7\textwidth]{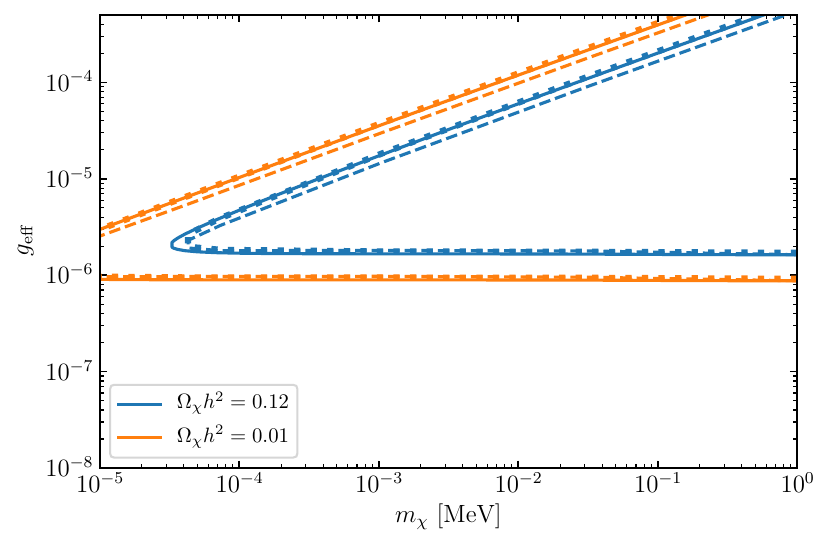}\caption{Required DM-neutrino interaction strength to produce the observed
		DM relic abundance or part of it. The solid curves represent results
		obtained by the model-independent approach---solving Eq.~\eqref{eq:-8}
		with the parametrization of $\langle\sigma v\rangle$ in Eq.~\eqref{eq:sigmav-formal}.
		Dashed and dotted (almost overlap with the solid) curves represent
		results obtained by solving the Boltzmann equations for the specific
		$s$- and $t$-channel models, including the complete set of relevant
		reactions---see Eqs.~\eqref{eq:s-1}-\eqref{eq:t-1}. \label{fig:DM-abundance}}
\end{figure}

By solving the Boltzmann equation \eqref{eq:-8} with the parametrization
of $\langle\sigma v\rangle$ in Eq.~\eqref{eq:sigmav-formal} numerically,
we obtain the final value of $\overline{n}_{\chi}$, which should
remain constant until today and can be converted to $\Omega_{\chi}h^{2}$
as follows:
\begin{equation}
	\Omega_{\chi}h^{2}=0.12\times\left.\frac{2m_{\chi}n_{\chi}}{\rho_{\text{DM},0}}\right|_{{\rm today}},\label{eq:-19}
\end{equation}
where $\rho_{{\rm DM},0}=9.74\times10^{-12}\ \text{eV}^{4}$ denotes
today's DM energy density. Note that the factor of $2$ comes from the degree of freedom of both $\chi$ and anti-$\chi$.

We then scan the parameter space to determine the required DM-neutrino
interaction strength to produce $\Omega_{\chi}h^{2}=0.12$. There
are three free parameters in our model-independent framework, namely
$m_{\chi}$, $\langle\sigma v\rangle_{0}$ and $\Lambda$. We find
that as long as $\Lambda/m_{\chi}$ is an ${\cal O}(1)$ quantity,
the result is not sensitive to the specific value of $\Lambda$. Hence
we fix $\Lambda=m_{\chi}/\sqrt{3}$ and focus on varying $m_{\chi}$
and $\langle\sigma v\rangle_{0}$. For the convenience of comparison with
specific models, we use the following dimensionless $g_{{\rm eff}}$
instead of $\langle\sigma v\rangle_{0}$:
\begin{equation}
	g_{{\rm eff}}\equiv\left(32\pi m_{\chi}^{2}\langle\sigma v\rangle_{0}\right)^{1/4}.\label{eq:-29}
\end{equation}
According to Eq.~\eqref{eq:-7}, $g_{{\rm eff}}$ corresponds to $\sqrt{g_{\nu}g_{\chi}}$
or $y_{\chi}$ in the $s$- and $t$-channel models, respectively.

In Fig.~\ref{fig:DM-abundance}, the blue solid curve is obtained
by requiring that $(m_{\chi},\ g_{{\rm eff}})$ leads to the observed
DM relic abundance. In addition to that, we also consider the scenario
that $\chi$ only accounts for part of the observed amount of DM\footnote{As we will show later, $\chi$ behaves as warm dark matter in the
	freeze-in and part of the freeze-out regimes. A mixture of cold and
	warm DM is a favored scenario when considering the {\it small-scale
			problems} of  cold DM. }. Hence we plot an orange solid curve corresponding to $\Omega_{\chi}h^{2}=0.01$.

The upper and lower branches of these curves correspond to the freeze-out
and freeze-in regimes, respectively. We find that the two branches
can be well fitted by the following expression:
\begin{equation}
	g_{{\rm eff}}\approx\left(\frac{\Omega_{\chi}h^{2}}{0.12}\right)^{\mp\frac{1}{4}}\times\begin{cases}
		1.6\times10^{-5}\left(\frac{m_{\chi}}{{\rm keV}}\right)^{0.53} & \text{freeze-out regime} \\
		1.8\times10^{-6}                                               & \text{freeze-in regime}
	\end{cases}\thinspace,\label{eq:-30}
\end{equation}
where $\mp$ takes minus and positive signs for the freeze-out and
freeze-in regimes, respectively.

\subsection{Model-specific analyses\label{subsec:Model-specific-analyses}}

When considering specific models such as the two in Eqs.~\eqref{eq:s}
and \eqref{eq:t}, the DM abundance may also be affected by the presence
of the mediator. For instance, in the $t$-channel model, the mediator
$\phi$ (which we assume is a complex scalar with a dark charge)
can be produced via $\nu\overline{\nu}\to\phi\phi^{*}$ and subsequently
decay to DM via $\phi\to\chi\nu$. In addition,  a $\phi\phi^{*}$
pair can annihilate to $\chi\overline{\chi}$. For the $s$-channel
model, there are also a few additional processes that need to be taken
into account. Overall, there are four processes in each of the two
models that can directly or indirectly affect the evolution of DM:
\begin{align}
	s\text{-channel model}:\  & \nu\overline{\nu}\leftrightarrow\phi\thinspace,\ \nu\overline{\nu}\leftrightarrow2\phi\thinspace,\ \nu\overline{\nu}\leftrightarrow\chi\overline{\chi}\thinspace,\ \chi\overline{\chi}\leftrightarrow2\phi\thinspace,\label{eq:s-1}     \\
	t\text{-channel model}:\  & \phi\leftrightarrow\chi\nu\thinspace,\ \nu\overline{\nu}\leftrightarrow\phi\phi^{*}\thinspace,\ \nu\overline{\nu}\leftrightarrow\chi\overline{\chi}\thinspace,\ \phi\phi^{*}\leftrightarrow\chi\overline{\chi}\thinspace.\label{eq:t-1}
\end{align}

In order to fully take these processes into account, we implement
a set of Boltzmann equations for the three coupled species, $\nu$,
$\chi$, and $\phi$, and solve the equations numerically. The implementation
is straightforward, with more details given in Appendix~\ref{sec:three-coupled}.

We add the results to Fig.~\ref{fig:DM-abundance}, presenting them
in dashed and dotted curves for the $s$- and $t$-channel models
respectively. The mediator mass is set at $m_{\phi}=5$ eV ($s$-channel)
and $m_{\phi}=1.2m_{\chi}$ ($t$-channel) in our calculation.

As is shown in Fig.~\ref{fig:DM-abundance}, the results obtained
for specific models are approximately the same as the one obtained
in the model-independent approach. For different models, the results
in terms of $g_{{\rm eff}}$ may vary by $\sim20\%$. This meets our
expectation since the contribution of addition production channels,
if comparable to that of $\nu\overline{\nu}\leftrightarrow\chi\overline{\chi}$,
can be absorbed by increasing or decreasing $\langle\sigma v\rangle_{0}$
by roughly a factor of two, corresponding to a variation of $g_{{\rm eff}}$
by $2^{1/4}-1\approx20\%$.

It is possible, however, by tuning the couplings and masses such that
one of those additional channels dominates significantly over $\nu\overline{\nu}\leftrightarrow\chi\overline{\chi}$.
For example, if $g_{\chi}\gg g_{\nu}$ in the $s$-channel model,
the relic abundance of $\chi$ in the freeze-out regime would mainly
depend on $\chi\overline{\chi}\leftrightarrow2\phi$, assuming $\phi$
is light. If $m_{\phi}$ is heavier than $2m_{\chi}$, the dominant
way of producing $\chi$ would be $\nu\overline{\nu}\to\phi$ followed
by $\phi\to\chi\overline{\chi}$. Such possibilities are beyond the
scope of our current framework which aims at a generic analysis based
on the minimal set of parameters.

\section{Cosmological consequences and constraints}
\label{sec:consequence}

\subsection{$N_{{\rm eff}}$ constraints}
\label{sec:Neff}

As was mentioned in Sec.~\ref{sec:introduction}, the conversion between
$\nu$ and $\chi$ could significantly modify the cosmological observable
$N_{{\rm eff}}$. The effect of $\chi$ on $N_{{\rm eff}}$ is two-fold:


(i) Before neutrino decoupling, some $\chi$ particles may have already been produced from the thermal bath. Although the process $\nu\overline{\nu}\leftrightarrow\chi\overline{\chi}$ cannot change $N_{\rm eff}$ when neutrinos are still in thermal equilibrium with the photon-electron plasma,  a significant amount of entropy stored in the $\chi$ sector will later be released into the neutrino sector after neutrino decoupling and increase $N_{\rm eff}$.

(ii) After neutrino decoupling, the reaction $\nu\overline{\nu}\to\chi\overline{\chi}$
consumes decoupled neutrinos and stores a significant amount of entropy into
the $\chi$ sector. Although a portion of the entropy (depending on
how large $R_{\Lambda}$ is) will be returned to $\nu$ at a later
phase, the evolution after neutrino decoupling overall leads to a
decrease of $N_{{\rm eff}}$.

\begin{figure}
	\centering
	\includegraphics[width=0.8\linewidth]{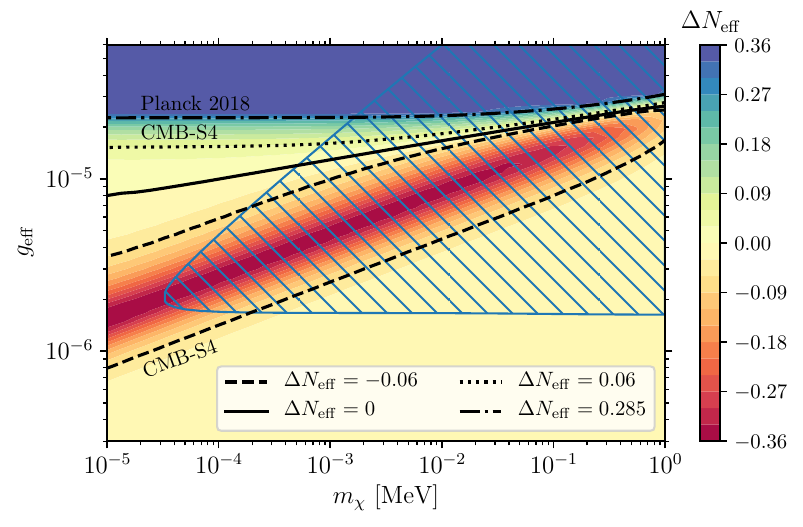}
	\caption{The effect of $\nu\overline{\nu}\leftrightarrow\chi\overline{\chi}$
	on $\Delta N_{{\rm eff}}$. The region below or above the black solid
	line leads to $\Delta N_{{\rm eff}}<0$ or $>0$, respectively. The
	hatched region causes an overproduction of DM ($\Omega_{\chi}h^{2}>0.12$).
	}
	\label{fig:Neff}
\end{figure}

Therefore, the deviation of $N_{{\rm eff}}$ from its standard value
$N_{{\rm eff}}^{{\rm st}}$, defined as
\begin{equation}
	\Delta N_{{\rm eff}}=N_{{\rm eff}}-N_{{\rm eff}}^{{\rm st}}\thinspace,\label{eq:-31}
\end{equation}
can be positive or negative, depending on whether (i) or (ii) is dominant.
Based on the arguments in (i) and (ii),  $\Delta N_{{\rm eff}}$ should
be  positive in the freeze-out
regime if $g_{{\rm eff}}$ is sufficiently large, or negative in the freeze-in regime.

We compute $\Delta N_{{\rm eff}}$ by numerically solving the Boltzmann
equations of the $\nu$-$\chi$ coupled sector and extracting the
final abundance of neutrinos after the freeze-out/in process has completed.
The result is presented in Fig.~\ref{fig:Neff}, which shows that
indeed the resulting $\Delta N_{{\rm eff}}$ is positive for large
$g_{{\rm eff}}$ and it turns negative when $g_{{\rm eff}}$ is below
$\sim10^{-5}$ (the black solid line). The negative $\Delta N_{{\rm eff}}$
can drop to as low as $-0.36$. This could be an interesting observational
consequence of our light DM scenario, since many new physics modifications
of $N_{{\rm eff}}$ often leads to positive $\Delta N_{{\rm eff}}$.

Currently the most precise measurement of $N_{{\rm eff}}$ comes from
Planck 2018~\cite{Planck:2018vyg}, $N_{{\rm eff}}=2.99\pm0.17$.
Subtracting the standard value $N_{{\rm eff}}^{{\rm st}}\approx3.045$~\cite{deSalas:2016ztq,EscuderoAbenza:2020cmq,Akita:2020szl,Bennett:2020zkv,Cielo:2023bqp}
from it, we obtain
\begin{equation}
	-0.395<\Delta N_{{\rm eff}}<0.285\thinspace,\ (2\sigma\ \text{C.L.})\thinspace.\label{eq:-32}
\end{equation}
We plot the upper bound as a dash-dotted curve in Fig.~\ref{fig:Neff}.
The lower bound, $-0.395$, is not sufficiently constraining to be
presented here. Future CMB experiments like Simons Observatory (SO)~\cite{SimonsObservatory:2019qwx,SimonsObservatory:2018koc},
CMB-S4~\cite{Abazajian:2019eic,CMB-S4:2016ple}, and CMB-HD~\cite{CMB-HD:2022bsz}
will significantly improve the measurement of $\Delta N_{{\rm eff}}$,
reaching the sensitivity of $0.1$, $0.06$, and $0.028$ at 2$\sigma$
C.L., respectively. This would allow the negative values of $\Delta N_{{\rm eff}}$
(e.g.~the part encompassed by the dashed curve in Fig.~\ref{fig:Neff})
to be probed.

\subsection{Lyman-$\alpha$ constraint}
\label{sec:lyman}

\begin{figure}
	\centering

	\includegraphics[width=0.7\textwidth]{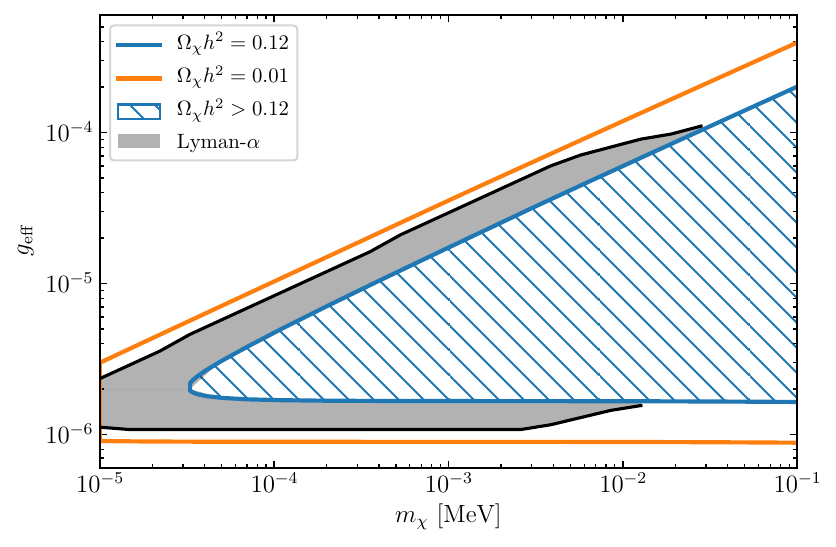}\caption{The Lyman-$\alpha$ constraint. The gray region would lead to an overlarge
	modification of the matter power spectrum at $\sim0.1$ Mpc scales
	and has been ruled out by Lyman-$\alpha$ observations. The hatched
	region causes an overproduction of DM ($\Omega_{\chi}h^{2}>0.12$).
	\label{fig:ly-alpha}}
\end{figure}

Light DM with relatively high velocity could lead to significant modifications
of the matter power spectrum at small ($\sim0.1$ Mpc) scales probed
by Lyman-$\alpha$ observations~\cite{Viel:2005qj,Viel:2013fqw,Baur:2017stq,Irsic:2017ixq,Palanque-Delabrouille:2019iyz,Garzilli:2019qki}. Therefore, Lyman-$\alpha$ observations
impose important constraints on our framework. While a dedicated study
on the Lyman-$\alpha$ constraints can be rather involved, there are
simplified approaches that allow for moderately accurate estimates
of the Lyman-$\alpha$ constraints --- see e.g.~\cite{Heeck:2017xbu,Decant:2021mhj,Coy:2021ann}.
Here we derive the Lyman-$\alpha$ constraints by means of the free
stream length, which is computed by
\begin{equation}
	\lambda_{{\rm FS}}=a_0\int_{t_{{\rm dec}}}^{t_{0}}\frac{v(t)}{a(t)}dt\thinspace,\label{eq:-34}
\end{equation}
where $v$ is the velocity of a DM particle, $t_{{\rm dec}}$ denotes
the time when DM becomes kinetically decoupled and starts free streaming,
$t_{0}$ denotes the time of today, and $a_0=a(t_0)$. Below we set $a_0=1$.

When the DM particle starts free streaming, its momentum scales as
$p\propto a^{-1}$, implying that the velocity $v=p/E_{p}=p/\sqrt{m_{\chi}^{2}+p^{2}}$
decreases as follows:
\begin{equation}
	v=\frac{p_{{\rm dec}}a_{{\rm dec}}/a}{\sqrt{m_{\chi}^{2}+p_{{\rm dec}}^{2}a_{{\rm dec}}^{2}/a^{2}}}\thinspace,\label{eq:-33}
\end{equation}
where $a_{{\rm dec}}$ and $p_{{\rm dec}}$ denote the values of $a$
and $p$ at kinetic decoupling.

The scale factor varies differently in the radiation-dominated and
matter-dominated eras:
\begin{equation}
	a=a_{\text{eq}}\left(\frac{t}{t_{\text{eq}}}\right)^{\gamma}\thinspace,\ \ \gamma=\begin{cases}
		1/2 & \text{radiation-dominated} \\
		2/3 & \text{matter-dominated}
	\end{cases}\thinspace,\label{eq:-35}
\end{equation}
where $a_{{\rm eq}}$ and $t_{{\rm eq}}$ denote the values of $a$
and $t$ at {\it matter-radiation equality}\footnote{Using the Planck 2018 measurements of $\Lambda$CDM model parameters~\cite{Planck:2018vyg,ParticleDataGroup:2022pth}, we obtain $a_{\rm eq}=2.9243\times 10^{-4}$, corresponding to red-shift $1+z=3419.63$ and $T_{\gamma}=0.803$ eV.}.

Substituting Eqs.~\eqref{eq:-33} and \eqref{eq:-35} into Eq.~\eqref{eq:-34},
we obtain
\begin{align}
	\lambda_{{\rm FS}}     & =\lambda_{\text{FS,RD}}+\lambda_{\text{FS,MD}}\thinspace,\label{eq:-36}                                                         \\
	\lambda_{\text{FS,RD}} & =2t_{{\rm eq}}\frac{\sinh^{-1}\eta_{{\rm eq}}-\sinh^{-1}\eta_{{\rm dec}}}{a_{{\rm eq}}\eta_{{\rm eq}}}\thinspace,\label{eq:-37} \\
	\lambda_{\text{FS,MD}} & =\frac{3t_{{\rm eq}}}{a_{{\rm eq}}\eta_{{\rm eq}}}\thinspace,\label{eq:-38}
\end{align}
where $\eta\equiv m_{\chi}/p$, with $\eta_{{\rm eq,dec}}=\eta(t=t_{{\rm eq,dec}})$.
Note that since $\sinh^{-1}\eta\approx\eta$ at $\eta\ll1$, if $\eta_{{\rm dec}}=m_{\chi}/p_{{\rm dec}}$
is sufficiently small (corresponding to sufficiently early decoupling),
then $\sinh^{-1}\eta_{{\rm dec}}$ in Eq.~\eqref{eq:-37} can be omitted,
resulting a free-streaming length almost independent of $\eta_{{\rm dec}}$.
Eq.~\eqref{eq:-38} is derived based on the non-relativistic assumption,
as the $\chi$ particle should already be non-relativistic at $t=t_{{\rm eq}}$.
In Eq.~\eqref{eq:-37}, we do not require it to be non-relativistic.
In fact, it is possible that $\chi$ transitions from relativistic
to non-relativistic regimes within the range covered by Eq.~\eqref{eq:-37}.
If we replace Eq.~\eqref{eq:-33} with a step function that equals
to $1$ or $v_{{\rm eq}}a_{{\rm eq}}/a$ in the relativistic and non-relativistic
regimes respectively, the above calculation reproduces Eq.~(38) in
Ref.~\cite{Coy:2021ann}. As we have checked, the numerical difference
between these two approaches is only around $10\%$.

To estimate the mean value of $\lambda_{{\rm FS}}$, we need to compute
the mean value of $p$:
\begin{equation}
	\langle p\rangle\equiv\frac{\int pf_{\chi}(p)d^{3}p}{\int f_{\chi}(p)d^{3}p}\thinspace.\label{eq:-40}
\end{equation}
When $f_{\chi}$ is in kinetic equilibrium, Eq.~\eqref{eq:-40} gives
\begin{equation}
	\frac{\langle p\rangle}{T}=\frac{2e^{-x}(x^{2}+3x+3)}{x^{2}K_{2}(x)}\approx\begin{cases}
		2\sqrt{\frac{2}{\pi}x}\approx1.6\sqrt{x} & \text{non-relativistic} \\
		3                                        & \text{relativistic}
	\end{cases}\thinspace.\label{eq:-41}
\end{equation}

We use Eq.~\eqref{eq:-41} to obtained the value of $p_{{\rm dec}}$
which is then used to computed the mean value of $\lambda_{{\rm FS}}$.
By matching the obtained $\lambda_{\text{FS}}$ with the free-streaming
length derived from Fig.~6 of Ref.~\cite{Baur:2017stq} (see also~\cite{Heeck:2017xbu}
for further interpretations), we can recast the Lyman-$\alpha$ constraint
reported in Ref.~\cite{Baur:2017stq} to the constraint on our model.
The result is presented as the gray region in Fig.~\ref{fig:ly-alpha}.

\subsection{Distortion of C$\nu$B}
\label{sec:CnuB}

As mentioned in Sec.~\ref{sec:introduction}, after neutrinos have kinetically
decoupled from $\chi$, the process $\chi\overline{\chi}\to\nu\overline{\nu}$
which continues until today (though at a very low rate) may lead to
a significant distortion of $f_{\nu}$. The resulting distortion of
$f_{\nu}$ cannot be washed out if it is produced after neutrinos
have started free-streaming.

\begin{figure}
	\centering

	\includegraphics[width=0.8\textwidth]{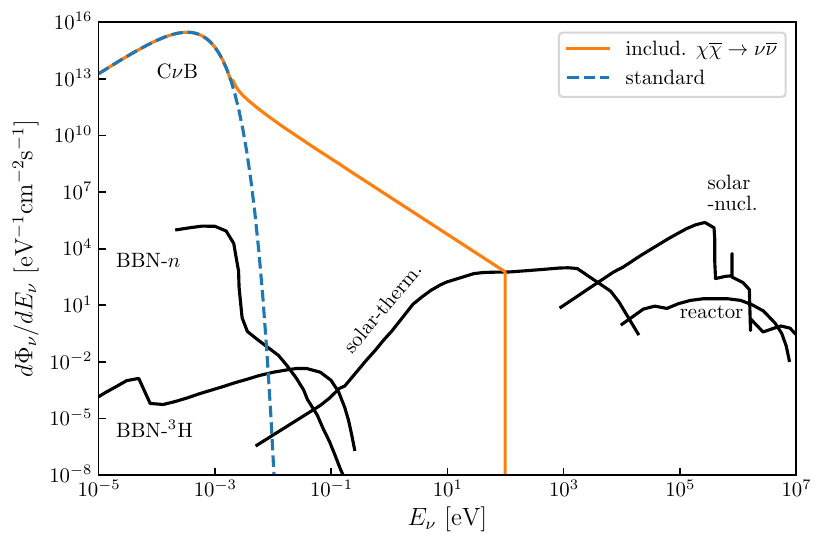}\caption{The impact of DM annihilation on C$\nu$B. The black curves represent
		other known neutrino fluxes at low energies, including those from
		BBN-$n$ and BBN-$^{3}{\rm H}$ decays, solar thermal (solar-therm.)
		and nuclear (solar-nucl.) reactions and reactors, taken from~\cite{Vitagliano:2019yzm}.
		\label{fig:CNB}}
\end{figure}

Before quantitative calculations, we shall clarify the difference
between the following two quantities:
\begin{equation}
	\tau_{\nu}=\frac{1}{\langle\sigma v\rangle_{\nu\chi\to\nu\chi}n_{\chi}}\thinspace,\ \ \tau_{\chi}=\frac{1}{\langle\sigma v\rangle_{\nu\chi\to\nu\chi}n_{\nu}}\thinspace,\label{eq:-44}
\end{equation}
where $\langle\sigma v\rangle_{\nu\chi\to\nu\chi}$ denotes the thermally-averaged
cross section of $\nu\chi\to\nu\chi$. Physically, we interpret $\tau_{\nu}$
as the mean free time (or mean free path, which is equivalent for
relativistic species) of a neutrino scattering off a $\chi$ particle,
and $\tau_{\chi}$ as the mean free time of a $\chi$ particle scattering
off a neutrino. Obviously, when $n_{\nu}\gg n_{\chi}$, we have $\tau_{\nu}\gg\tau_{\chi}$,
implying that neutrinos can start free-streaming much earlier than
$\chi$. This is also quite intuitive---when $\chi$ particles are
very scarce in the universe, it is much more difficult for a neutrino
to find a $\chi$ particle to collide with than a $\chi$ particle
to find a neutrino to collide with.

Therefore, when the terminology {\it kinetic decoupling} is used,
one should distinguish between the kinetic decoupling of $\nu$ from
$\chi$ and that of $\chi$ from $\nu$. The former happens when $\tau_{\nu}$
starts to exceed $H^{-1}$, meaning that the time it would take for
a neutrino to find a $\chi$ particle is longer than the Hubble time.
Likewise, the kinetic decoupling of $\chi$ from $\nu$ occurs when
$\tau_{\chi}\gtrsim H^{-1}$.

To compute the distortion, we adopt the Boltzmann equation of $f_{\nu}$
and focus only on the period when $\tau_{\nu}\gtrsim H^{-1}$. The
Boltzmann equation reads:
\begin{equation}
	\left[\frac{\partial}{\partial t}-Hp\frac{\partial}{\partial p}\right]f_{\nu}(p,t)=C^{(f)}\thinspace,\label{eq:-42}
\end{equation}
where the collision term $C^{(f)}$ is given as follows:
\begin{equation}
	C^{(f)}=\frac{2\pi^{2}}{m_{\chi}^{2}}n_{\chi}^{2}\langle\sigma v\rangle_{0}\delta(p-m_{\chi})\thinspace.\label{eq:-49}
\end{equation}
Note that when integrating over $p$, it reproduces the collision
term for $n_{\nu}$:
\begin{equation}
	\int C^{(f)}\frac{d^{3}p}{(2\pi)^{3}}=n_{\chi}^{2}\langle\sigma v\rangle_{0}\thinspace.\label{eq:-50}
\end{equation}

The Boltzmann equation can be analytically solved by means of variable
transformation $(t,\ p)\to(a,\ x_{p})$ with $x_{p}\equiv p/T$---see
Appendix~B in Ref.~\cite{Li:2022bpp} for further details.
Using this method, the correction to $f_{\nu}$ caused by $\chi\overline{\chi}\to\nu\overline{\nu}$ at  $\tau_{\nu}\gtrsim H^{-1}$ can be computed by
\begin{equation}
	\delta f_{\nu}=\int_{a_{\nu\text{FS}}}^{a_{0}}\frac{C^{(f)}(a,\ x_{p})}{Ha}da=\frac{2\pi^{2}n_{\chi}^{2}}{H_{m}m_{\chi}^{2}}\langle\sigma v\rangle_{0}x_{p}^{2}\Theta\thinspace,\label{eq:-51}
\end{equation}
where $a_{\nu\text{FS}}$ denotes the scale factor when $\nu$ starts
free-streaming, and $\Theta=1$ if $x_{p}^{\min}<x_{p}<x_{p}^{\max}$
or $0$ if $x_{p}$ is not in this range. The minimum and maximum
of $x_{p}$ are determined by requiring that $\int_{a_{\nu\text{FS}}}^{a_{0}}\delta(x_{p}T-m_{\chi})da$ is nonzero.
Note that here and also in Eq.~\eqref{eq:-51}, $x_{p}$ is treated
as an $a$-independent variable, while $T\propto a^{-1}$ is $a$-dependent.
The maximum of $x_{p}$ corresponds to DM annihilation today which
produces $\nu$ without red-shift, while earlier production of $\nu$
is subjected to higher red-shift. Hence the maximum is given by $x_{p}^{\max}=m_{\chi}/T_{0}$
where $T_{0}$ denotes the value of $T$ today (approximately the
neutrino temperature today),  and the minimum is given by $x_{p}^{\min}=a_{\nu\text{FS}}x_{p}^{\max}$
which is a red-shifted value of $x_{p}^{\max}$.

Here we should clarify a subtle difference between the number density
$n_{\chi}$ in Eq.~\eqref{eq:-50} and the $n_{\chi}$ in Eq.~\eqref{eq:-51}.
The former varies in time, while the latter is no longer time-dependent
after integrating over $a$,
essentially equivalent to integrating over time.
The former $n_{\chi}$ is actually a function of $a$ while
the latter $n_{\chi}$ should be a function of $x_{p}$. By checking
how the Dirac delta function is integrated out, we can see that the
latter $n_{\chi}$ is the number density of $\chi$ at $x_{p}=ax_{p}^{\max}$.
Consequently, the dependence of the former $n_{\chi}$ on $a$ transforms
into the dependence of the latter $n_{\chi}$ on $x_{p}$ via $n_{\chi}=n_{\chi}|_{a\to x_{p}/x_{p}^{\max}}$.

Let us consider a specific benchmark as a case study. It is based
on the $t$-channel model, with the following input parameters:
\begin{equation}
	y=10^{-5}\thinspace,\ m_{\chi}=0.1\ {\rm keV}\thinspace,\ m_{\phi}/m_{\chi}=1.2\thinspace.\label{eq:-46}
\end{equation}

The thermally-averaged cross section of $\nu\chi\to\nu\chi$ in the
$t$-channel model is given by
\begin{equation}
	\langle\sigma v\rangle_{\nu\chi\to\nu\chi}\approx\frac{3T^{2}y^{4}}{\pi\left(m_{\phi}^{2}-m_{\chi}^{2}\right)^{2}}\thinspace,\ \ (\text{for }T\ll m_{\phi}-m_{\chi})\thinspace,\label{eq:-45}
\end{equation}
where we have made non-relativistic approximation of $\chi$ and taken
the low-$T$ limit.

Using Eq.~\eqref{eq:-45} and Eq.~\eqref{eq:-44}, we obtain
\begin{equation}
	\frac{\tau_{\nu}}{H^{-1}}\approx\frac{1.0\ \text{eV}^{3}}{n_{\chi}}\left(\frac{m_{\chi}}{0.1\ {\rm keV}}\right)^{4}\cdot\left(\frac{10^{-5}}{y}\right)^{4}.\label{eq:-47}
\end{equation}
This ratio would exceed $1$ when $n_{\chi}$ is below $1\text{eV}^{3}$
for the given benchmark. According to the numerical solution, this
occurs when $x=m_{\chi}/T$ arrives at
\begin{equation}
	x_{{\rm \nu FS}}\approx 12.6\thinspace.\label{eq:-48}
\end{equation}
It is also straightforward to obtain the freeze-out value for this
benchmark, $x_{{\rm f.o}}\approx5.0$. By comparing the two values,
one can see that neutrinos start free-streaming later than freeze-out.

It is straightforward to evaluate the quantities in Eq.~\eqref{eq:-51}
for the above benchmark and obtain the distortion of C$\nu$B, which
is presented in Fig.~\ref{fig:CNB}. Here the $y$-axis is presented
in terms of the neutrino differential flux $d\Phi_{\nu}/dE_{\nu}$,
which for the isotropic and homogeneous C$\nu$B is related to $f_{\nu}$
by $d\Phi_{\nu}/dE_{\nu}=p^{2}f_{\nu}/(2\pi^{2})$. According to Eq.~\eqref{eq:-51},
the flux generated by DM annihilation is $d\Phi_{\nu}/dE_{\nu}\propto p^{2}n_{\chi}^{2}x_{p}^{2}\propto E_{\nu}^{-2}$.
Therefore, the high-energy tail of the C$\nu$B spectrum is altered
from exponential suppression to $E_{\nu}^{-2}$, substantially enhancing
the C$\nu$B flux at high energies. Note that there is a sharp cutoff at  $E_\nu=m_{\chi}$, corresponding to neutrinos produced from today's DM annihilation. In principle, the cutoff is not infinitely sharp since DM particles today still have small yet non-vanishing velocities and the velocity dispersion should slightly soften the cutoff.

It is particularly interesting to note that the enhanced C$\nu$B
tail would entirely change the landscape of existing neutrino spectra
in the sub-eV to keV range, in which the only known neutrino sources in
addition to C$\nu$B are neutron ($n$) and tritium ($^{3}{\rm H}$)
decays during BBN and solar thermal neutrinos~\cite{Vitagliano:2019yzm}. Their
fluxes are well below the orange curve for $E_{\nu}\leq m_{\chi}$.
If future experiments are capable to detect neutrinos within the eV-keV
range, the altered C$\nu$B tail would be the most predominant signal
in this energy range. In particular, for C$\nu$B detection experiments
such as PTOLEMY~\cite{PTOLEMY:2018jst,PTOLEMY:2019hkd}, in which the primary
challenge lies in sub-eV energy measurements~\cite{Cheipesh:2021fmg,PTOLEMY:2022ldz},
the distorted C$\nu$B in our framework offers a technically more
accessible scenario to probe.

\section{Summary and conclusions}
\label{sec:summary}

A dark sector interacting with the SM via the neutrino portal is one of the most appealing scenarios discussed in the literature.
Such interactions can not only be responsible for the thermal production of dark matter (DM) but also leave cosmological imprints at various stages of the early universe.

In this paper, we systematically investigate a generic framework in which DM interacts with neutrinos, and the interaction strength can be effectively quantified by the thermally-averaged cross section $\langle \sigma v \rangle$. We propose a model-independent parametrization of $\langle \sigma v \rangle$ and demonstrate that it can accurately account for the production and annihilation rates of DM across a wide range of temperatures, applicable to both relativistic and non-relativistic DM particles.  By feeding the parametrized  $\langle \sigma v \rangle$ into the Boltzmann equation, we solve the Boltzmann equation and obtain the DM relic abundance for a broad range of parameter space. When specific models are considered, the results are approximately the same as those obtained in our model-independent approach, as shown in Figs.~\ref{fig:sigma-v} and \ref{fig:DM-abundance}. Moreover, our parametrized  $\langle \sigma v \rangle$ allows the thermal evolution of DM to be computed analytically, with the analytic result agreeing well with the numerical one.

The interplay between DM with neutrinos may result in interesting cosmological imprints on neutrinos. We show that $N_{\rm eff}$ can be modified significantly due to DM production and annihilation. A particularly noteworthy feature here is that large negative $\Delta N_{\rm eff}$ (e.g.~$-0.36$) is possible and lies slightly beyond the reach of the Planck 2018 measurement. Future experiments such as CMB-S4 are capable to probe this interesting deviation.
Furthermore, DM annihilation at late times after kinetic decoupling might distort the C$\nu$B energy spectrum, yielding a substantially enhanced C$\nu$B tail with the neutrino flux much higher than that from BBN isotope decays and solar thermal neutrinos---see Fig.~\ref{fig:CNB}. Such an enhancement would have great implications for future C$\nu$B detection experiments.

\begin{acknowledgments}
	We thank Laura Lopez-Honorez and Keisuke Harigaya for useful discussions.
	Fermilab is operated by Fermi Research
	Alliance, LLC under Contract No.~DE-AC02-07CH11359
	with the U.S. Department of Energy, Office of Science, Office of High Energy Physics.
	I.R.W. is supported by DOE distinguished scientist fellowship grant FNAL 22-33.
	I.R.W was supported by DOE-SC0010008 when at Rutgers at the early stage of this work.
	The work of X.J.X is supported in part by the National Natural Science
	Foundation of China under grant No.~12141501 and also by the CAS
	Project for Young Scientists in Basic Research (YSBR-099).
	X.J.X would also like to thank the Hangzhou Institute for Advanced Study (HIAS) for the hospitality during his visit when this work was performed in part.
\end{acknowledgments}

\appendix

\section{Collision terms and thermally-averaged cross sections\label{sec:Collision-terms} }

The Boltzmann equation in Eq.~\eqref{eq:-8} in terms of $\langle\sigma v\rangle$
is derived from the following form
\begin{equation}
	\frac{dn_{\chi}}{dt}+3Hn_{\chi}=-N_{\nu}C_{\chi\overline{\chi}\leftrightarrow\nu\overline{\nu}}\thinspace,\label{eq:-8-1}
\end{equation}
where
\begin{equation}
	C_{\chi\overline{\chi}\leftrightarrow\nu\overline{\nu}}\equiv\int\left[f_{\chi}(p_{1})f_{\overline{\chi}}(p_{2})-f_{\nu}(p_{3})f_{\overline{\nu}}(p_{4})\right]|{\cal M}|^{2}(2\pi\delta)^{4}d\Pi_{1}d\Pi_{2}d\Pi_{3}d\Pi_{4}\thinspace.\label{eq:-23}
\end{equation}
Due to kinetic equilibrium {[}see Eq.~\eqref{eq:-20}{]}, $\langle\sigma v\rangle$
can be equivalently given by the following four different forms:
\begin{align}
	\langle\sigma v\rangle & =n_{\chi}^{-2}\int f_{\chi}(p_{1})f_{\chi}(p_{2})|{\cal M}|^{2}(2\pi\delta)^{4}d\Pi_{1}d\Pi_{2}d\Pi_{3}d\Pi_{4}\label{eq:-24}                                                                                 \\
	                       & =\left(n_{\chi}^{{\rm eq}}\right)^{-2}\int f_{\chi}^{{\rm eq}}(p_{1})f_{\chi}^{{\rm eq}}(p_{2})|{\cal M}|^{2}(2\pi\delta)^{4}d\Pi_{1}d\Pi_{2}d\Pi_{3}d\Pi_{4}\label{eq:-25}                                   \\
	                       & =\left(n_{\chi}^{{\rm eq}}\right)^{-2}\int f_{\nu}^{{\rm eq}}(p_{3})f_{\nu}^{{\rm eq}}(p_{4})|{\cal M}|^{2}(2\pi\delta)^{4}d\Pi_{1}d\Pi_{2}d\Pi_{3}d\Pi_{4}\label{eq:-26}                                     \\
	                       & =\left(\frac{n_{\nu}^{{\rm eq}}}{n_{\chi}^{{\rm eq}}}\right)^{2}\frac{1}{n_{\nu}^{2}}\int f_{\nu}(p_{3})f_{\nu}(p_{4})|{\cal M}|^{2}(2\pi\delta)^{4}d\Pi_{1}d\Pi_{2}d\Pi_{3}d\Pi_{4}\thinspace.\label{eq:-27}
\end{align}
Note that from Eq.~\eqref{eq:-25} to Eq.~\eqref{eq:-26}, the $\delta$
function allows the replacement $f_{\chi}^{{\rm eq}}(p_{1})f_{\chi}^{{\rm eq}}(p_{2})=e^{-(E_{1}+E_{2})/T}\to f_{\nu}^{{\rm eq}}(p_{3})f_{\nu}^{{\rm eq}}(p_{4})=e^{-(E_{3}+E_{4})/T}$.

Applying Eqs.~\eqref{eq:-24} and \eqref{eq:-27} to Eq.~\eqref{eq:-23},
we obtain
\begin{equation}
	\frac{dn_{\chi}}{dt}+3Hn_{\chi}=-N_{\nu}\langle\sigma v\rangle\left[n_{\chi}^{2}-\left(\frac{n_{\chi}^{{\rm eq}}}{n_{\nu}^{{\rm eq}}}\right)^{2}n_{\nu}^{2}\right],\label{eq:-28}
\end{equation}
which is essentially Eq.~\eqref{eq:-8}.

For other two-to-two processes we have encountered in this work such
as $2\phi\to2\chi$, one can obtain similar equations, independent
of whether the initial and final states are massless or not.

The collision term $C_{\chi\overline{\chi}\leftrightarrow\nu\overline{\nu}}$
in Eq.~\eqref{eq:-23} can be split into two terms
\begin{equation}
	C_{\chi\overline{\chi}\leftrightarrow\nu\overline{\nu}}=C_{\chi\overline{\chi}\to\nu\overline{\nu}}-C_{\nu\overline{\nu}\to\chi\overline{\chi}}\thinspace,\label{eq:C-diff}
\end{equation}
where $C_{\chi\overline{\chi}\to\nu\overline{\nu}}$ and $C_{\nu\overline{\nu}\to\chi\overline{\chi}}$
correspond to the integral in Eq.~\eqref{eq:-23} with $[f_{\chi}f_{\overline{\chi}}-f_{\nu}f_{\overline{\nu}}]$
replaced by $f_{\chi}f_{\overline{\chi}}$ and $f_{\nu}f_{\overline{\nu}}$,
respectively. From Eq.~\eqref{eq:-10}, we have
\begin{equation}
	\langle\sigma v\rangle=\frac{1}{n_{\chi}^{2}}C_{\chi\overline{\chi}\to\nu\overline{\nu}}=\left(\frac{n_{\nu}^{{\rm eq}}}{n_{\chi}^{{\rm eq}}}\right)^{2}\frac{1}{n_{\nu}^{2}}C_{\nu\overline{\nu}\to\chi\overline{\chi}}\thinspace.\label{eq:-52}
\end{equation}

The collision terms $C_{\chi\overline{\chi}\to\nu\overline{\nu}}$
and $C_{\nu\overline{\nu}\to\chi\overline{\chi}}$ can be computed
by integrating the cross sections of their respective processes\,---\,see
Ref.~\cite{Gondolo:1990dk}. Specifically, assuming $f_{\nu(\overline{\nu})}(p)=e^{-p/T}$
or $f_{\chi(\overline{\chi})}(p)=e^{-E_{p}/T}$ with $E_{p}=\sqrt{m_{\chi}^{2}+p^{2}}$,
we have
\begin{align}
	C_{\nu\overline{\nu}\to\chi\overline{\chi}} & =\frac{T}{32\pi^{4}}\int_{4m_{\chi}^{2}}^{\infty}s^{3/2}\sigma_{\nu\overline{\nu}\to\chi\overline{\chi}}K_{1}\left(\frac{s^{1/2}}{T}\right)ds\thinspace,\label{eq:-53}                             \\
	C_{\chi\overline{\chi}\to\nu\overline{\nu}} & =\frac{T}{32\pi^{4}}\int_{4m_{\chi}^{2}}^{\infty}s^{1/2}\left(s-4m_{\chi}^{2}\right)\sigma_{\chi\overline{\chi}\to\nu\overline{\nu}}K_{1}\left(\frac{s^{1/2}}{T}\right)ds\thinspace.\label{eq:-54}
\end{align}
If $\nu$ or $\chi$ are not in chemical equilibrium but in kinetic
equilibrium, Eqs.~\eqref{eq:-53} and \eqref{eq:-54} are changed
by a factor of $(n_{\nu}/n_{\nu}^{{\rm eq}})^{2}$ or $(n_{\chi}/n_{\chi}^{{\rm eq}})^{2}$.
Either Eq.~\eqref{eq:-53} or Eq.~\eqref{eq:-54} can be used to compute
$\langle\sigma v\rangle$ in Eq.~\eqref{eq:-52}, and would lead to
the same result.

For the specific $s$- and $t$-channel models introduced in Eqs.~\eqref{eq:s}
and \eqref{eq:t}, the cross sections are given by\footnote{Computations for the cross sections are cross-checked in this work under the help of \textsc{FeynCalc 9.3.0}~\cite{Mertig:1990an,Shtabovenko:2016sxi,Shtabovenko:2020gxv} and \textsc{Package-X}~\cite{Patel:2015tea,Patel:2016fam}.}~\cite{Hufnagel:2021pso}
\begin{align}
	\sigma_{\nu\overline{\nu}\to\chi\overline{\chi}}^{(s)} & =\frac{g_{\chi}^{2}g_{\nu}^{2}}{12\pi}\frac{s-m_{\chi}^{2}}{\left(s-m_{\phi}^{2}\right)^{2}}\Delta\thinspace,\label{eq:-59}                                                                                      \\
	\sigma_{\nu\overline{\nu}\to\chi\overline{\chi}}^{(t)} & =\frac{y_{\chi}^{4}}{16\pi s^{2}}\left[\frac{sm_{\phi}^{2}+2\delta^{4}}{sm_{\phi}^{2}+\delta^{4}}s\Delta+2\delta^{2}\log\left(\frac{1-\Delta+2\delta^{2}/s}{1+\Delta+2\delta^{2}/s}\right)\right],\label{eq:-60}
\end{align}
with
\begin{equation}
	\Delta\equiv\sqrt{1-\frac{4m_{\chi}^{2}}{s}}\thinspace,\ \ \delta^{2}\equiv m_{\phi}^{2}-m_{\chi}^{2}\thinspace.\label{eq:-61}
\end{equation}

For analytical calculations, it is useful to mention the following
limits
\begin{equation}
	\lim_{\delta\to0}\sigma_{\nu\overline{\nu}\to\chi\overline{\chi}}^{(t)}=\frac{y_{\chi}^{4}}{16\pi s}\Delta\thinspace,\ \ \lim_{m_{\phi}\to0}\sigma_{\nu\overline{\nu}\to\chi\overline{\chi}}^{(s)}=\frac{g_{\chi}^{2}g_{\nu}^{2}}{12\pi s}\left(1-\frac{m_{\chi}^{2}}{s}\right)\Delta\thinspace.\label{eq:-6}
\end{equation}
With the cross sections in Eq.~\eqref{eq:-6},  $\langle\sigma v\rangle$
can be computed analytically in the low-$T$ and high-$T$ limits.

In the low-$T$ limit, assuming $x=m_{\chi}/T\gg1$, we obtain
\begin{align}
	\langle\sigma v\rangle^{(t)} & =\frac{y_{\chi}^{4}}{32\pi m_{\chi}^{2}}\left[1-3x^{-1}+{\cal O}\left(x^{-2}\right)\right],\label{eq:-55}                      \\
	\langle\sigma v\rangle^{(s)} & =\frac{g_{\chi}^{2}g_{\nu}^{2}}{32\pi m_{\chi}^{2}}\left[1-\frac{5}{2}x^{-1}+{\cal O}\left(x^{-2}\right)\right].\label{eq:-56}
\end{align}

In the high-$T$ limit, assuming $x=m_{\chi}/T\ll1$, we obtain
\begin{align}
	\langle\sigma v\rangle^{(t)} & =\frac{y_{\chi}^{4}}{128\pi T^{2}}+{\cal O}\left(x^{2}\right),\label{eq:-57}           \\
	\langle\sigma v\rangle^{(s)} & =\frac{g_{\chi}^{2}g_{\nu}^{2}}{96\pi T^{2}}+{\cal O}\left(x^{2}\right).\label{eq:-58}
\end{align}

For general values of $T$ (e.g.~for the solid curves in Fig.~\ref{fig:sigma-v}),
we compute the integral in Eq.~\eqref{eq:-53} numerically to obtain
$C_{\nu\overline{\nu}\to\chi\overline{\chi}}$ and hence $\langle\sigma v\rangle$.

\section{Useful approximations in the freeze-out regime\label{sec:f-o-approx}}

The Boltzmann equation in the freeze-out regime can be reformulated
into the following form which is essentially a Riccati equation:
\begin{equation}
	y'(x)=-\frac{y(x)^{2}-y_{0}(x)^{2}}{x^{2}}\thinspace,\label{eq:y}
\end{equation}
with
\begin{equation}
	y_{0}(x)=R\frac{x^{2}K_{2}(x)}{2}\approx R\times\begin{cases}
		1                                            & \ \text{for }x\ll1 \\
		\frac{1}{2}\sqrt{\frac{\pi}{2}}x^{3/2}e^{-x} & \ \text{for }x\gg1
	\end{cases}\thinspace.\label{eq:}
\end{equation}

\begin{figure}
	\centering

	\includegraphics[width=0.7\textwidth]{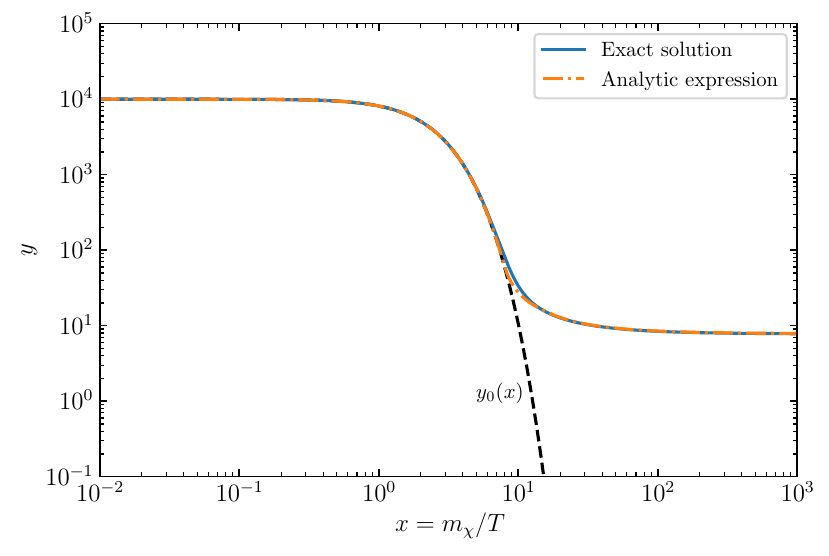}\caption{Comparison of the analytical expression in Eq.~\eqref{eq:-4} with
	the exact solution obtained by numerically solving Eq.~\eqref{eq:y}
	with $R=10^{4}$. \label{fig:fo}}
\end{figure}

Eq.~\eqref{eq:y} in general cannot be solved analytically. But in
the limit of vanishing $y_{0}$ it has the following solution
\begin{equation}
	y=\frac{x}{x/y_{\infty}-1}\thinspace,\ \ \ (y_{0}\to0)\thinspace,\label{eq:-2}
\end{equation}
where $y_{\infty}$ denotes the value of $y$ at $x\to\infty$. The
solution is divergent at $x=y_{\infty}$, which implies that it is
invalid when $x$ is below $y_{\infty}$. Therefore, the freeze-out
value of $x$ is approximately given by
\begin{equation}
	x_{{\rm f.o}}\approx y_{\infty}\thinspace.\label{eq:-3}
\end{equation}
For $x\ll x_{{\rm f.o}}$, $\chi$ should be in equilibrium, implying
that the solution is approximately given by the equilibrium value,
$y(x)\approx y_{0}(x)$. Combining this and Eq.~\eqref{eq:-2}, we
obtain the following analytical solution:
\begin{equation}
	y=\begin{cases}
		\frac{x}{x/x_{{\rm f.o}}-1}                                     & x>2x_{{\rm f.o}}                     \\
		y_{0}(x)                                                        & x<x_{{\rm f.o}}                      \\
		\frac{x\thinspace y_{{\rm f.o}}}{x(\xi-1)+x_{{\rm f.o}}(2-\xi)} & x\in[x_{{\rm f.o}},\ 2x_{{\rm f.o}}]
	\end{cases}\thinspace,\label{eq:-4}
\end{equation}
where $y_{{\rm f.o}}\equiv y_{0}(x_{{\rm f.o}})$ and $\xi\equiv y_{{\rm f.o}}/x_{{\rm f.o}}$.
Here we assume a transition interval $[x_{{\rm f.o}},\ 2x_{{\rm f.o}}]$.
Within this interval, the solution is obtained by assuming an expression
of the form $x/(ax+b)$ with $a$ and $b$ determined by the continuity
of the solution.

In Fig.~\ref{fig:fo}, we show that Eq.~\eqref{eq:-4} as an approximate
solution is very close to the exact one. However, we should note that
in the use of Eq.~\eqref{eq:-4} one has to determine $x_{{\rm f.o}}$
accurately. This could be obtained from the widely used criteria that
$x_{{\rm f.o}}$ corresponds to $n_{\chi}\langle\sigma v\rangle_{0}\sim H$.
A more accurate result can be obtained by numerically solving Eq.~\eqref{eq:y}
with various given values of $R$ and then performing a polynomial
(in terms of $\log_{10}R$) fit to the results. In this way, we find
\begin{equation}
	x_{{\rm f.o}}\approx0.048+1.73\log_{10}R+0.051\left(\log_{10}R\right)^{2}\thinspace,\label{eq:-5}
\end{equation}
which serves as a formula for computing $x_{{\rm f.o}}$ with precision
at the percent level.

\section{\label{sec:three-coupled}Solving Boltzmann equations for three coupled
  species}

In Sec.~\ref{subsec:Model-specific-analyses}, we computed the DM
relic abundance for the specific $s$- and $t$-channel models. This
requires solving the Boltzmann equations for three coupled species:
\begin{equation}
	\frac{dn_{i}}{dt}+3Hn_{i}=C_{{\rm prod.}}^{(n_{i})}-C_{{\rm depl.}}^{(n_{i})}\thinspace,\ \ (i=\nu,\ \chi,\ \phi)\thinspace.\label{eq:-62}
\end{equation}
where  $C_{{\rm prod.}}^{(n_{i})}$ and $C_{{\rm depl.}}^{(n_{i})}$
denote the collision terms accounting for the production and the depletion
of $n_{i}$. In our numerical code, we solve the Boltzmann equations
by reformulating them into the following form
\begin{equation}
	\frac{d\left(n_{i}a^{3}\right)}{da}=\frac{a^{2}}{H}\left[C_{{\rm prod.}}^{(n_{i})}-C_{{\rm depl.}}^{(n_{i})}\right].\label{eq:-43}
\end{equation}

According to Eq.~\eqref{eq:s-1}, the collision terms for the $s$-channel
model include the following contributions:
\begin{align}
	C_{{\rm prod.}}^{(n_{\nu})}  & =N_{\phi}C_{\phi\to\nu\overline{\nu}}+N_{\phi}C_{2\phi\to\nu\overline{\nu}}+C_{\chi\overline{\chi}\to\nu\overline{\nu}}\thinspace,\label{eq:-63} \\
	C_{{\rm depl.}}^{(n_{\nu})}  & =N_{\phi}C_{\nu\overline{\nu}\to\phi}+N_{\phi}C_{\nu\overline{\nu}\to2\phi}+C_{\nu\overline{\nu}\to\chi\overline{\chi}}\thinspace,\label{eq:-64} \\
	\nonumber                                                                                                                                                                       \\
	C_{{\rm prod.}}^{(n_{\chi})} & =N_{\phi}C_{2\phi\to\chi\overline{\chi}}+N_{\nu}C_{\nu\overline{\nu}\to\chi\overline{\chi}}\thinspace,\label{eq:-65}                             \\
	C_{{\rm depl.}}^{(n_{\chi})} & =N_{\phi}C_{\chi\overline{\chi}\to2\phi}+N_{\nu}C_{\chi\overline{\chi}\to\nu\overline{\nu}}\thinspace,\label{eq:-66}                             \\
	\nonumber                                                                                                                                                                       \\
	C_{{\rm prod.}}^{(n_{\phi})} & =N_{\nu}C_{\phi\to\nu\overline{\nu}}+2N_{\nu}C_{\nu\overline{\nu}\to2\phi}+2C_{\chi\overline{\chi}\to2\phi}\thinspace,\label{eq:-67}             \\
	C_{{\rm depl.}}^{(n_{\phi})} & =N_{\nu}C_{\nu\overline{\nu}\to\phi}+2N_{\nu}C_{2\phi\to\nu\overline{\nu}}+2C_{2\phi\to\chi\overline{\chi}}\thinspace,\label{eq:-68}
\end{align}
where $N_{\phi}=3$ accounts for the three polarization modes of the
vector $\phi$.

It is worth mentioning a few consistency checks for the coefficients
in Eqs.~\eqref{eq:-63}-\eqref{eq:-68}. When combining the Boltzmann
equations of two species together, there should be some kind of cancellations
among the right-hand sides of the Boltzmann equations.  For instance,
if only $\phi\leftrightarrow\nu\overline{\nu}$ is present, then we
expect that the total number of $\phi$ and $\nu$ in a comoving volume
is conserved, i.e. $d\left[(N_{\nu}n_{\nu}+N_{\phi}n_{\phi})a^{3}\right]/da=0$.
Indeed, one can see that the combination {[}Eq.~\eqref{eq:-63}-Eq.~\eqref{eq:-64}{]}$\times N_{\nu}+${[}Eq.~\eqref{eq:-67}$-$Eq.~\eqref{eq:-68}{]}$\times N_{\phi}$
vanishes if only $\phi\leftrightarrow\nu\overline{\nu}$ is present.
Similar consistency checks also apply to other terms. Specifically,
when only $2\phi\leftrightarrow\nu\overline{\nu}$, $\chi\overline{\chi}\leftrightarrow\nu\overline{\nu}$
or $\chi\overline{\chi}\leftrightarrow2\phi$ is present, then $(2N_{\nu}n_{\nu}+N_{\phi}n_{\phi})a^{3}$,
$(N_{\nu}n_{\nu}+n_{\chi})a^{3}$ or $(2n_{\chi}+N_{\phi}n_{\phi})a^{3}$
is conserved, respectively.

For the $t$-channel model, Eqs.~\eqref{eq:-63}-\eqref{eq:-68} should
be changed to the following forms:
\begin{align}
	C_{{\rm prod.}}^{(n_{\nu})}  & =C_{\phi\to\chi\nu}+C_{\phi\phi^{*}\to\nu\overline{\nu}}+C_{\chi\overline{\chi}\to\nu\overline{\nu}}\thinspace,\label{eq:-69}          \\
	C_{{\rm depl.}}^{(n_{\nu})}  & =C_{\chi\nu\to\phi}+C_{\nu\overline{\nu}\to\phi\phi^{*}}+C_{\nu\overline{\nu}\to\chi\overline{\chi}}\thinspace,\label{eq:-70}          \\
	\nonumber                                                                                                                                                             \\
	C_{{\rm prod.}}^{(n_{\chi})} & =C_{\phi\phi^{*}\to\chi\overline{\chi}}+N_{\nu}C_{\nu\overline{\nu}\to\chi\overline{\chi}}\thinspace,\label{eq:-71}                    \\
	C_{{\rm depl.}}^{(n_{\chi})} & =C_{\chi\overline{\chi}\to\phi\phi^{*}}+N_{\nu}C_{\chi\overline{\chi}\to\nu\overline{\nu}}\thinspace,\label{eq:-72}                    \\
	\nonumber                                                                                                                                                             \\
	C_{{\rm prod.}}^{(n_{Z'})}   & =N_{\nu}C_{\chi\nu\to\phi}+N_{\nu}C_{\nu\overline{\nu}\to\phi\phi^{*}}+C_{\chi\overline{\chi}\to\phi\phi^{*}}\thinspace,\label{eq:-73} \\
	C_{{\rm depl.}}^{(n_{Z'})}   & =N_{\nu}C_{\phi\to\chi\nu}+N_{\nu}C_{\phi\phi^{*}\to\nu\overline{\nu}}+C_{\phi\phi^{*}\to\chi\overline{\chi}}\thinspace.\label{eq:-74}
\end{align}
Again, one can check that there are similar cancellations among the
collision terms to guarantee that some combinations of particle numbers
are conserved.

\bibliographystyle{JHEP}
\bibliography{ref}

\end{document}